\documentclass[aps,prl,twocolumn,superscriptaddress,preprintnumbers,showpacs,showkeys]{revtex4}

\usepackage[dvips]{graphicx}
\usepackage{epsbox,color}
\usepackage{amsmath}
\usepackage{float}

\begin{document}

\preprint{YITP-11-47}

\title{Charmonium spectra at finite temperature from QCD sum rules with the maximum entropy method}


\author{Philipp Gubler}
\email{phil@th.phys.titech.ac.jp}
\affiliation{Department of Physics, H-27, Tokyo Institute of Technology, Meguro, Tokyo 152-8551, Japan}

\author{Kenji Morita}
\affiliation{Yukawa Institute for Theoretical Physics, Kyoto University, Kyoto 606-8502, Japan}
\affiliation{GSI, Helmholzzentrum f$\ddot{u}$r Schwerionenforschung, Planckstrasse 1, D-64291 Darmstadt, Germany}

\author{Makoto Oka}
\affiliation{Department of Physics, H-27, Tokyo Institute of Technology, Meguro, Tokyo 152-8551, Japan}

\date{\today}

\begin{abstract}
Charmonia spectral functions at finite temperature are studied using QCD sum rules in combination with 
the maximum entropy method. 
This approach enables us to directly obtain the spectral function from the sum rules, 
without having to introduce any specific assumption about its functional form. 
As a result, it is found that while $J/\psi$ and $\eta_c$ manifest themselves as significant 
peaks in the spectral function below the deconfinement temperature $T_c$, they quickly dissolve into the 
continuum and almost completely disappear at temperatures between 1.0 $T_c$ and 1.1 $T_c$.       

\end{abstract}

\pacs{12.38.Lg, 12.38.Mh, 14.40.Lb}
\keywords{Charmonium, QCD sum rules, QCD at finite temperature}

\maketitle

Since QCD was established to be the theory of strong interactions, charmonium 
has often been used as a suitable probe of its dynamics, owing to the fact that in this system 
both perturbative and nonperturbative aspects of QCD play equally important roles \cite{Novikov}. 
The behavior of charmonia in a hot or dense medium has also attracted much interest, 
as it was suggested some time ago, that in the color-deconfined medium with a temperature 
above $T_c$ 
charmonia will dissolve due to the color Debye screening, and thus serve as a signal 
for the formation of quark-gluon plasma \cite{Matsui}.

Testing these early suggestions from first principles of QCD has become feasible only 
recently, as new developments in lattice QCD have made it possible to access the 
charmonium spectral functions with the help of the maximum entropy method (MEM) 
\cite{Asakawa,Datta,Umeda,Jakovac}. 
These studies found that the lowest charmonium states ($J/\psi$ and $\eta_c$) survive up to 
temperatures as high as $\sim$ 1.5 $T_c$ or higher.

Besides lattice QCD, the method of QCD sum rules \cite{Shifman} provides another tool for investigating 
the properties of hadrons at finite temperature \cite{Bochkarev,Hatsuda}. Using this 
approach various charmonium channels were studied recently \cite{Morita1,Morita3}, and evidence for a 
considerable change of the spectral functions just above $T_c$ was found. 
To specify the nature of this change is the major goal of this study. For 
this task we employ MEM, which is applicable to QCD sum rules \cite{Gubler} and has 
the advantage that one does not have to introduce any strong assumption about the 
functional form of the spectral function, such as the ``pole + continuum" ansatz, 
which is often used in QCD sum rule studies.

Let us first recapitulate 
what sort of information QCD sum rules can provide on the charmonium spectral function at 
finite temperature \cite{Bochkarev,Hatsuda}. 
One considers the time-ordered correlator at finite temperature 
\begin{equation} 
\Pi^{\mathrm{J}}(q) = i \displaystyle \int d^4x e^{iqx} 
\langle T [j^{\mathrm{J}}(x) j^{\mathrm{J}}(0) ] \rangle_T, 
\label{eq:correlator} 
\end{equation}
where $j^{\mathrm{J}}(x)$ stands for $\bar{c} \gamma_{\mu} c(x)$ and $\bar{c} \gamma_{5} c(x)$ in the vector ($\mathrm{V}$)
and pseudoscalar ($\mathrm{PS}$) channel, respectively. The expectation value $\langle \mathcal{O} \rangle_T$ is defined 
as $\langle \mathcal{O} \rangle_T \equiv \mathrm{Tr}( e^{-H/T} \mathcal{O} ) / \mathrm{Tr}( e^{-H/T} )$. 
Throughout this work, we will set the spatial momentum of the charmonium system relative to the thermal medium to be 
$\textbf{0}$; thus, $q^{\mu} = (\omega, \textbf{0})$. In this circumstance, there is only one independent component 
in the correlator of the vector channel. 
In what follows, we will use the dimensionless functions  
$\tilde{\Pi}^{\mathrm{V}}(q^2) \equiv \Pi^{\mu,\mathrm{V}}_{\mu}(q)/(-3q^2)$ and 
$\tilde{\Pi}^{\mathrm{PS}}(q^2) \equiv \Pi^{\mathrm{PS}}(q)/q^2$ for the analysis. 

Going to the deep Euclidean region $q^2 \equiv -Q^2 \ll 0$, one can calculate 
the correlation functions using the 
operator product expansion (OPE), giving an expansion in local operators $O_n$ with increasing mass dimension $n$: 
$\tilde{\Pi}^{\mathrm{J}}(q^2) = \sum_n C^{\mathrm{J}}_n(q^2) \langle O_n \rangle_T$. As was first discussed in \cite{Hatsuda}, 
as long as the temperature $T$ lies below the separation scale of the OPE, which is of the order of $\sim 1$ GeV, 
all the temperature effects can be included into the expectation values of the local operators $\langle O_n \rangle_T$, while 
the Wilson coefficients $C^{\mathrm{J}}_n(q^2)$ are independent of $T$. 
Furthermore, to improve the convergence of the OPE and suppressing the influence of high energy states onto the sum rule, we 
apply the Borel transform to the correlator, leading to 
the final result of the OPE for $\nu \equiv 4m_c^2/M^2$, $M$ being the Borel mass: 
\begin{equation}
\begin{split}
\mathcal{M}^{\mathrm{J}}(\nu) = & e^{-\nu}A^{\mathrm{J}}(\nu)[1 + \alpha_s(\nu) a^{\mathrm{J}}(\nu) 
+ b^{\mathrm{J}}(\nu) \phi_b(T) \\
&+ c^{\mathrm{J}}(\nu) \phi_c(T) +  d^{\mathrm{J}}(\nu) \phi_d(T)].
\end{split}
\label{eq:OPE}
\end{equation} 
The first two terms in 
Eq.(\ref{eq:OPE}) are the leading order perturbative term and its first order 
$\alpha_s$ correction. The third and fourth terms contain the scalar and 
twist-2 gluon condensates of mass dimension 4: 
$\phi_b(T) = \frac{4\pi^2}{9(4m_c^2)^2} G_0$ 
and $\phi_c(T) = \frac{4\pi^2}{3(4m_c^2)^2} G_2$, where 
$G_0 = \langle \frac{\alpha_s}{\pi} G^a_{\mu\nu} G^{a\mu\nu}\rangle_T$ and $G_2$ is 
defined as $\langle \frac{\alpha_s}{\pi} G^{a\mu\sigma} G^{a\nu}_{\sigma}\rangle_T 
= (u^{\mu}u^{\nu} - \frac{1}{4}g^{\mu\nu})G_2$, $u^{\mu}$ being the four velocity 
of the medium. 
For the detailed expressions of the Wilson coefficients of these 
terms, see \cite{Morita3}. 
To evaluate the 
possible influence of higher order contributions, we include one more term, 
which is proportional to the scalar gluon 
condensate of dimension 6, $\phi_d(T) = \frac{1}{(4m_c^2)^3}\langle g^3 f^{abc} G^{a\nu}_{\mu} G^{b\lambda}_{\nu} 
G^{c\mu}_{\lambda} \rangle_T$. The Wilson coefficient of this term can be found in \cite{Marrow}. 

The correlator can also be expressed by a dispersion relation, 
in terms of the spectral function $\rho^{\mathrm{J}}(\omega)$ 
of the channel specified by the operator $j^{\mathrm{J}}(x)$. After the Borel 
transform one obtains 
\begin{equation}
\mathcal{M}^{\mathrm{J}}(\nu) = \displaystyle \int_0^{\infty}dx^2  e^{-x^2 \nu} \rho^{\mathrm{J}}(2m_c x).
\label{eq:dispersion}
\end{equation} 
Equating Eqs.(\ref{eq:OPE}) and (\ref{eq:dispersion}) then gives the final form of the sum rules.
In the vector channel, an additional constant term contributes to Eq.(\ref{eq:OPE}), which originates 
from a pole at $\omega = 0$ in $\rho^{\mathrm{V}}(\omega)$ \cite{Bochkarev}. As this so-called scattering term 
considerably complicates the analysis, we eliminate it by 
taking the derivative of Eqs.(\ref{eq:OPE}) and (\ref{eq:dispersion}) with respect to $\nu$ and analyze only the 
resulting derivative sum rule in this channel. For a discussion on the validity of 
this procedure in the heavy quark sum rules, see \cite{Morita4}. 

The usual strategy of analyzing QCD sum rules is to make some reasonable 
assumptions on the functional form of the spectral function, 
and then extract information 
on the lowest lying peak from Eqs.(\ref{eq:OPE}) and (\ref{eq:dispersion}). This method, however, has 
several shortcomings. First of all, the widely used ``pole + continuum" ansatz, which certainly works 
well at $T=0$, may not be appropriate at temperatures above $T_c$, where the lowest lying 
state is expected to be modified and eventually melt into the $c$-$\bar{c}$ continuum, which could become  
the dominant contribution. Furthermore, it is not always possible to 
unambiguously fit a specific ansatz to the OPE results, because of the occurrence of equally 
valid solutions. Such a situation arose in 
\cite{Morita1,Morita3}, where it was not possible to determine a unique solution 
for the used parametrization of the spectral function. 
To handle these problems, we propose to use MEM, which allows us to extract the spectral function from 
Eqs.(\ref{eq:OPE}) and (\ref{eq:dispersion}) without prejudice on its functional form. Moreover, it can 
be proven that this 
method provides a unique solution for the spectral function \cite{Asakawa2}. 

Let us now briefly 
summarize the basic ideas of MEM, which helps us to carry out the task of inverting 
the integral of Eq.(\ref{eq:dispersion}). This is, however, an ill-posed problem as we have only 
access to $\mathcal{M}^{\mathrm{J}}(\nu)$ in the region of $\nu$ where the OPE shows sufficient 
convergence and, furthermore, have only information on $\mathcal{M}^{\mathrm{J}}(\nu)$ with limited 
precision due to the uncertainties of the values of $m_c$, $\alpha_s$ and the various condensates. 
Nonetheless, using Bayesian probability theory, 
MEM makes it possible to at least obtain the most probable form of the spectral function $\rho(\omega)$. 
To do this, one defines a probability $P[\rho|\mathcal{M}\mathcal{I}]$ for $\rho$ to have a specific form 
given $\mathcal{M}$ and additional information $\mathcal{I}$ on $\rho$ such as positivity and asymptotic 
values. Using Bayes' theorem, this probability can be rewritten as $P[\rho|\mathcal{M}\mathcal{I}] \propto
P[\mathcal{M}|\rho\mathcal{I}] P[\rho|\mathcal{I}] = e^{- L + \alpha S}$, where $e^{-L}$ is the likelihood function, 
used in standard $\chi^2$ methods. 
$e^{\alpha S}$ stands for the prior probability, given by the 
parameter $\alpha$ and the Shannon-Jaynes entropy, 
\begin{equation}
S = \displaystyle \int_0^{\infty} d\omega \Bigr[ \rho(\omega) - m(\omega) - \rho(\omega)
\log \Bigl( \frac{\rho(\omega)}{m(\omega)} \Bigr) \Bigl].
\label{eq:SJe}
\end{equation}
$m(\omega)$ is called the default model and can be used to incorporate prior knowledge on 
the asymptotic values of $\rho(\omega)$ into the analysis. Determining now the most probable $\rho(\omega)$ 
corresponds to the numerical problem of finding the maximum of the functional $- L + \alpha S$, 
for which we use the Bryan algorithm \cite{Bryan}. 
As a last step, the spectral function $\rho_{\alpha}(\omega)$ maximizing $- L + \alpha S$ for a fixed  
value of $\alpha$ is averaged using 
$\displaystyle \int d \alpha \rho_{\alpha}(\omega) P[\alpha |\mathcal{M}\mathcal{I}]$, which 
gives the final output of the MEM analysis. For more details see \cite{Asakawa2}, while specific issues 
concerning the application of this method to QCD sum rules are discussed in \cite{Gubler}.    

Next, we describe how 
the temperature dependencies of the gluonic condensates are determined. 
For the scalar and twist-2 gluon condensates with mass dimension 4, we follow the 
approach proposed in \cite{Morita1,Morita3}, where, in the quenched approximation, the energy momentum tensor, 
expressed using gluonic operators, was matched with the corresponding quantity written down in form of the energy 
density $\epsilon$ and the pressure $p$, leading to 
$G_0 = G^{\mathrm{vac}}_0 
- \frac{8}{11}\bigl[\epsilon(T) - 3p(T)\bigr]$ and 
$G_2 = -\frac{\alpha_s(T)}{\pi}\bigl[\epsilon(T) + p(T)\bigr]$ 
for the scalar and twist-2 gluon condensates. The functions 
$\epsilon(T)$, $p(T)$ and $\alpha_s(T)$ were then extracted from quenched 
lattice QCD data \cite{Boyd,Kaczmarek}. 
We will in this study use the same numerical values for the 
$T$ dependent part of $G_0$ and $G_2$ as in \cite{Morita3}. As is shown there, 
both $G_0$ and $G_2$ exhibit a sudden decrease in the vicinity of $T_c$. 

It was suggested in previous studies that the OPE could break down at 
temperatures above $T_c$ as higher dimensional operators may become non-negligible \cite{Morita3}. 
To investigate this possibility, we include 
the scalar gluonic condensate with dimension 6, $\langle g^3 f^{abc} G^{a\nu}_{\mu} G^{b\lambda}_{\nu} 
G^{c\mu}_{\lambda} \rangle$, about which much less is known. To our 
knowledge, at $T=0$, there exists only an estimate based on the dilute instanton gas model, 
giving 
$
\langle g^3 f^{abc} G^{a\nu}_{\mu} G^{b\lambda}_{\nu} 
G^{c\mu}_{\lambda} \rangle = \frac{48 \pi^2}{5\rho_c^2} \langle \frac{\alpha_s}{\pi} G^2 \rangle, 
$ 
where $\rho_c$ is a representative value for the instanton radius, 
for which we use the established value of 0.33 fm \cite{Schafer}. 
Assuming that the relation above also holds at finite temperature, and taking into account the
reduction of $\rho_c$ above $T_c$ \cite{Chu}, we, however, 
found that the dimension 6 term does not influence the behavior of the spectral 
function much in the temperature region investigated in this Letter. 
Therefore, we conclude that even though the relation obtained from the dilute instanton gas model 
can only be considered to be a crude estimate, as
long as it gives the correct order of magnitude, the contribution of the dimension 6 condensate 
is small and does not alter the results obtained in 
this study.

\begin{figure*}
\includegraphics[width=7.0cm,clip]{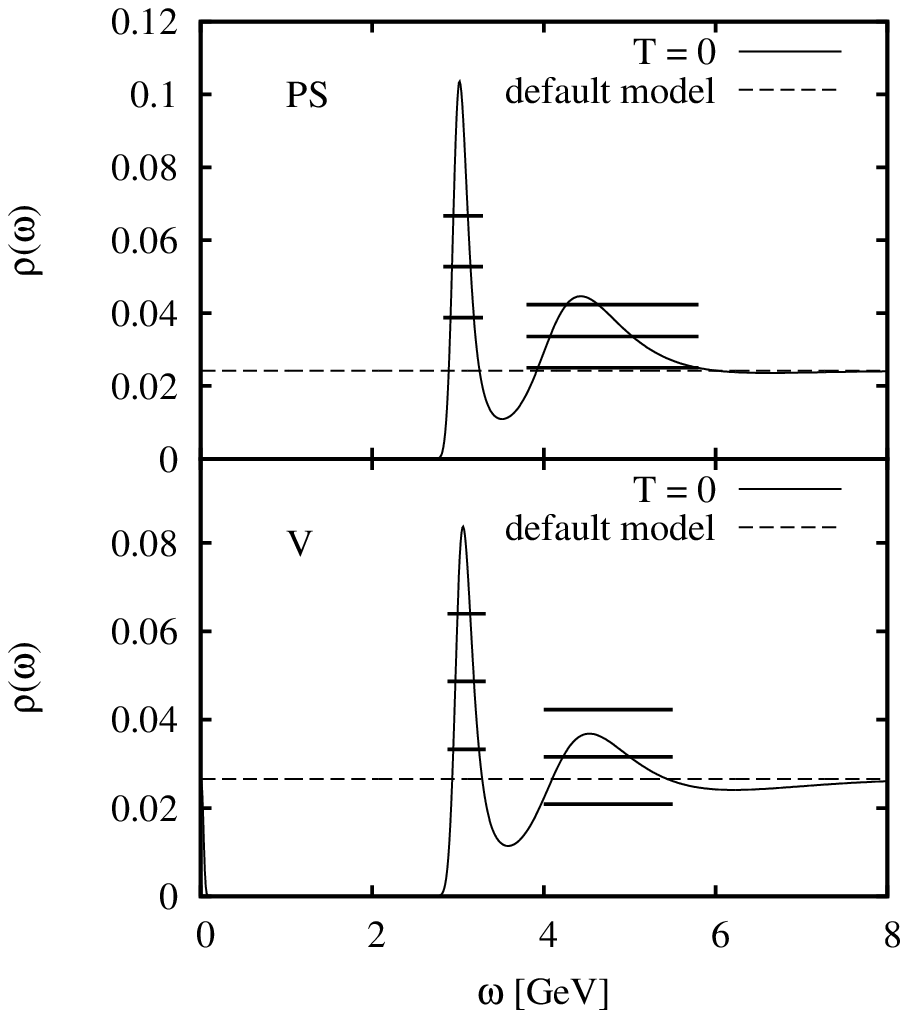}
\hspace{1.0cm}
\includegraphics[width=7.0cm,clip]{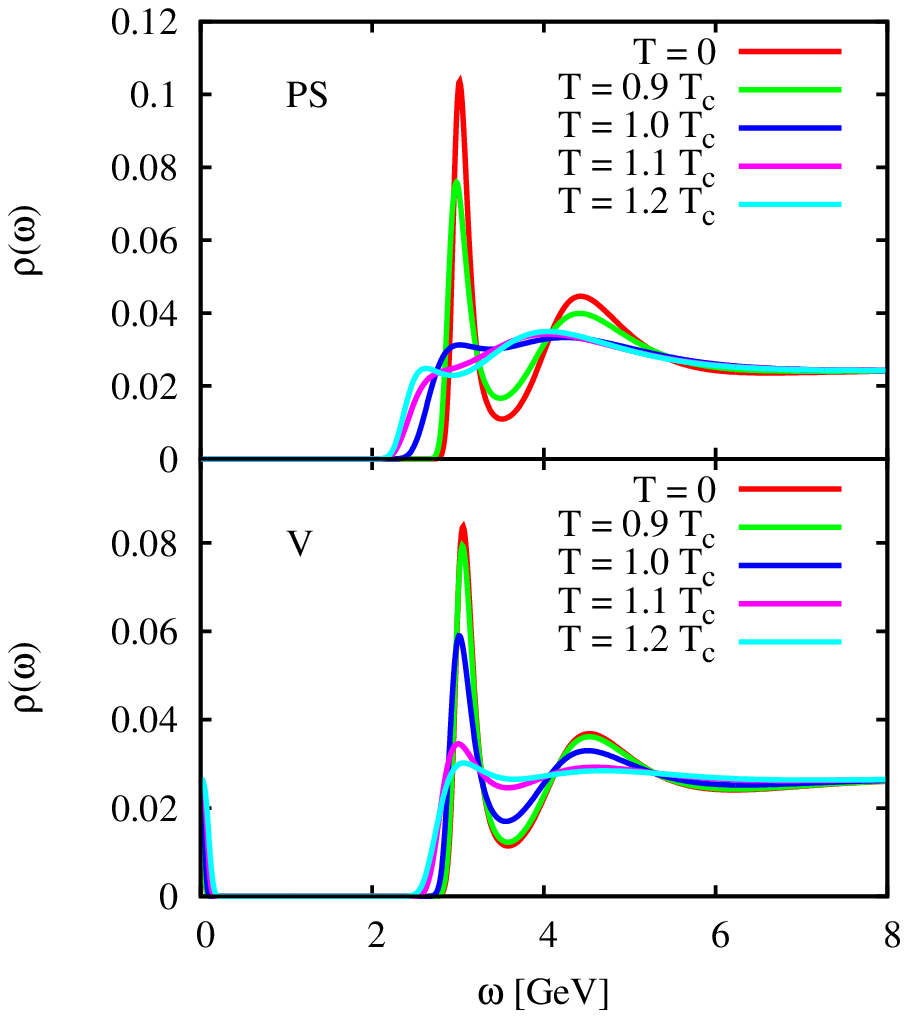}
\caption{On the left side, the spectral functions in the pseudoscalar (upper plot) and vector (lower plot) channel at $T=0$, with the 
errors of their averaged values are shown. 
The horizontal extent and position of the error bars indicates the region over which the average is taken. 
The dashed lines show the default model used in the MEM analysis. 
On the right side, the corresponding spectral functions are given at temperatures around $T_c$.}
\label{fig:zerofinite}
\end{figure*}
Let us now turn to the MEM analysis of Eqs.(\ref{eq:OPE}) and (\ref{eq:dispersion}). First, we investigate the 
spectral function at $T = 0$ in the vector and pseudoscalar channel. To determine the upper boundary of the region of $\nu$ 
to be analyzed, we employ the criterion that the dimension 6 term should be smaller than 20\% of the whole OPE expression 
of Eq.(\ref{eq:OPE}), which gives $\nu^{\mathrm{V}}_{\mathrm{max}} = 8.03$ in the vector and $\nu^{\mathrm{PS}}_{\mathrm{max}} = 7.29$ 
in the pseudoscalar channel. We keep these values fixed when going to finite temperature. In fact, 
in the temperature region around $T_c$, the relative contribution of the dimension 6 term at $\nu^{\mathrm{V},\mathrm{PS}}_{\mathrm{max}}$ 
is even smaller, 
namely, around 10\% or less. The lower boundary of $\nu$ is chosen to be $\nu^{\mathrm{V},\mathrm{PS}}_{\mathrm{min}} = 0.78$, 
corresponding to a Borel mass of $M = 3.0$ GeV. 
We have checked that the obtained 
spectral functions do not depend much on this choice. 
For the value of the charm quark mass $m_c$, we use a recent estimate giving $\overline{m}_c(m_c) = 1.277 \pm 0.026$ GeV \cite{Dehnadi}, 
for $\alpha_s$ we employ the newest world average $\alpha_s(M_Z) = 0.1184 \pm 0.0007$ \cite{Bethke}, 
while for the vacuum gluon condensate $G^{\mathrm{vac}}_0$ the standard value 
$G^{\mathrm{vac}}_0 = 0.012 \pm 0.0036$ $\mathrm{GeV}^4$ \cite{Shifman,Colangelo} is applied. 
For the default model 
$m(\omega)$, we use a constant matched to the perturbative value of the spectral function at high energy, as was done in 
similar studies using lattice QCD \cite{Asakawa}. 
 
The resulting spectral functions are given on the left side of Fig. \ref{fig:zerofinite}. 
We observe in both channels a clear ground state peak, corresponding to $\eta_c$ and $J/\psi$. 
The spectral functions also exhibit a second peak, which is, however, not statistically significant. These second 
peaks most likely reflect the existence of several excited states, which the MEM analysis is not able to resolve, 
quite similar to the situation encountered in lattice studies. 
Furthermore, it is seen that the spectral function of the vector channel approaches the default model in the region 
close to $\omega = 0$, which, however, should not be confused with a contribution of the scattering term. 
This behavior is an artifact caused by our usage of the derivative sum rule in this channel and should thus not be 
considered to be a physical effect. 
Numerically, the peak representing $\eta_c$ lies at $3.02$ GeV, while the one standing for $J/\psi$ is found at $3.06$ GeV. 
Thus, we see that the ground state in both channels 
reproduces the experimental value with a precision of the order of $50$ MeV. 
In the vector channel, the residue can be related to the electronic width of the corresponding resonance. 
We can obtain this residue from Fig. \ref{fig:zerofinite} simply by integrating the spectral function in the region of 
the peak, which gives 0.162 $\mathrm{GeV}^2$, which is in good agreement with 
the experimental value of 0.173 $\mathrm{GeV}^2$. 
On the other hand, we observe that the hyperfine splitting 
between $\eta_c$ and $J/\psi$ is underestimated. 
All these findings are in qualitative agreement with the results obtained in 
the conventional analysis of the charmonium sum rules \cite{Marrow}. 
    
Next, we increase the temperature according to Eq.(\ref{eq:OPE}). The resulting spectral 
functions are shown on the right side of Fig. \ref{fig:zerofinite} at temperatures between $0.9$ $T_c$ and $1.2$ $T_c$. 
It is seen in the figure that the behavior of the spectral functions changes abruptly in the 
vicinity of $T_c$. First, both ground state peaks experience a shift to lower energies of the order of 50 MeV,  
before dissolving quickly into the continuum above the critical temperature. 
Investigating the spectral functions in more detail, one observes that $\eta_c$ disappears already at 
$T=1.0$ $T_c$, while $J/\psi$ survives a bit longer, but also appears to be melted to a large degree at 
$T=1.1$ $T_c$. 
This sudden qualitative change of the spectral function mainly originates from the changes of the third 
and fourth terms in Eq.(\ref{eq:OPE}), which 
can be traced back to the rapid adjustment of the thermodynamic quantities 
$\epsilon(T)$ and $p(T)$ around $T_c$. 
It is reassuring to note that our results are consistent with the findings of \cite{Morita1,Morita3} 
in the sense that both observe a negative energy shift of the peaks around $T_c$. 
In these earlier works, it was, however, not possible to discuss the possible melting of the peaks because 
a relativistic Breit-Wigner form for the spectral function was assumed at all investigated temperatures. 

For obtaining firm conclusions, one has to test the reliability of the MEM procedure at finite temperature, 
where systematic effects decrease the reproducibility and resolution of the spectral function obtained from MEM. 
In lattice studies, this reduced reliability is primarily caused by the reduction of the data points 
in the imaginary time correlator, due to periodicity and the reduction of the maximal time extent. 
In the case of QCD sum rules, this problem does not exist, as Eq.(\ref{eq:OPE}) is given as 
a continuous function at any temperature and therefore the same number of data points can be used. Nevertheless, the reliability of 
the MEM technique is still reduced at finite temperature due to the uncertainties of the thermodynamic functions 
$\epsilon(T)$ and $p(T)$, whose contribution grows with temperature and therefore increases the error of Eq.(\ref{eq:OPE}). 
In order to confirm that the change of the spectral function in Fig. \ref{fig:zerofinite} is not an artifact, 
we reanalyze Eq.(\ref{eq:OPE}) at $T=0$, but use the errors of $T\neq0$ in the analysis. The results are 
then compared to the ones given in Fig. \ref{fig:zerofinite}, to investigate the net temperature effect on the 
spectral function. 
We find from this analysis 
that while the height of the peaks of the spectral functions at $T=0$ is indeed reduced because of the 
increased error, this effect is much smaller than the actual reduction of the peaks around $T_c$, 
seen on the right side of Fig. \ref{fig:zerofinite}. 
We therefore 
conclude that the disappearance of the peaks observed in Fig. \ref{fig:zerofinite} is a physical effect and is not 
induced by an artifact of the MEM analysis. 

In summary, we have extracted the spectral functions of the pseudoscalar and vector channel at both zero and finite 
temperature using a combined analysis of QCD sum rules and MEM. At $T=0$, the MEM technique is able to clearly resolve 
the lowest energy peaks, corresponding to the $\eta_c$ and $J/\psi$ resonances. The positions of both peaks agree with the 
experimental values with a precision of about 50 MeV. 
At finite temperature, we find that $\eta_c$ and $J/\psi$ melt quickly after the temperature is raised above the 
deconfinement temperature $T_c$, caused by the sudden change of the dimension-4, scalar and twist-2 gluon 
condensates in this temperature region. We have checked that this effect is not an artifact of the systematics of the 
MEM analysis. These results quantitatively disagree with the earlier findings of lattice studies which suggest that both 
$\eta_c$ and $J/\psi$ can survive at temperatures of up to 1.5 $T_c$ or higher. It, however, has to be mentioned 
that our results are in fact consistent with the latest lattice results \cite{Ding}, finding the peaks of 
$\eta_c$ and $J/\psi$ to be largely distorted between $0.73$ $T_c$ and $1.46$ $T_c$. 
It remains to be seen whether or not 
the two methods will converge to compatible conclusions when more accurate analyses will become available 
in the future.

\begin{acknowledgments}
This work was supported by KAKENHI under Contracts No. 22105503, No. 
19540275 and 
by YIPQS at the Yukawa Institute for Theoretical Physics. 
P.G. gratefully 
acknowledges the support by the Japan Society for the Promotion of Science for Young 
Scientists (Contract No. 21.8079). 
K.M. thanks FIAS for support. 
\end{acknowledgments}




\end{document}